# DIVO: a direct divertor configuration plasma start-up scheme aimed at tokamak fusion reactors

**Wenbin Liu[1], Jinping Qian[1,*], Long Zeng[2,*], Rui Ding[1,*], Yuehang Wang[1], Richard A. Pitts[3], Ernesto Lerche[4], D. B. King[5], Bin Zhang[1], Xianzu Gong[1], Zhengping Luo[1], Yao Huang[1], Binfu Gao[1], Dalong Chen[1], Pengfei Zhou[1], Fang Ding[1], Liqing Xu[1], Yanmin Duan[1], Chaowei Mai[6], Weixi Cao[1], Wei Gao[1], Shouxin Wang[1], Weiye Xu[1], Handong Xu[1], Bin Cao[1], Kedong Li[1], Zhe Gao[2], Ximan Li[2], Ling Zhang[1], Yunxin Cheng[1], Kangning Geng[1] and Hailing Zhao[1]**

[1] Institute of Plasma Physics, Hefei Institutes of Physical Science, Chinese Academy of Sciences, Hefei 230031, China
[2] Department of Engineering Physics, Tsinghua University, Beijing 100084, China
[3] ITER Organization, Route de Vinon-sur-Verdon, CS 90 046 13067, St. Paul Lez Durance Cedex, France
[4] Laboratory for Plasma Physics, LPP-ERM/KMS, B-1000, Brussels, Belgium
[5] United Kingdom Atomic Energy Authority, Culham Campus, Abingdon, Oxfordshire OX14 3DB, United Kingdom of Great Britain and Northern Ireland
[6] Guangdong Ocean University, Zhanjiang 524088, China

Email: jpqian@ipp.ac.cn; zenglong@tsinghua.edu.cn; rding@ipp.ac.cn

## Abstract

The tokamak, one of the mainstream approaches to magnetic confinement in nuclear fusion devices, will be the first approach to reach reactor class prototyping. Plasma start-up is expected to be more challenging in large superconducting next-step tokamaks such as ITER due to the low inductive toroidal electric field and potentially long durations in limiter configuration when plasma-wall interactions lead to easier impurity contamination of the confined plasma. Such limiter start-up is the standard way of early plasma formation in the vast majority of today's experimental devices and is planned on ITER. To mitigate these issues, a novel plasma start-up scheme with divertor geometry only (DIVO start-up) was recently developed for the first time on the EAST superconducting tokamak. The key breakthrough in this work is the use of a non-inductive, electron cyclotron wave generated plasma formed in a poloidal field configuration which is a superposition of a barrel-like geometry around the centre of the vacuum vessel and a mirror-like geometry across the whole vessel cross-section. Such a non-inductive discharge avoids the drawback of low inductive electric field which is unavoidable in large superconducting devices as a consequence of the relatively low voltages which can be applied to the magnet coils. In addition to the low toroidal electric field, the DIVO start-up scheme as demonstrated on EAST significantly reduces impurity sources compared to the conventional limiter phase approach, even in cases with relatively poorly conditioned plasma-facing surfaces. The robustness of the DIVO start-up has also been verified on EAST with varying prefill gas pressure, impurity level and inductive electric field. It thus offers a potential solution to one of the main issues facing routine successful start-up on large tokamak reactors.

# 1. Plasma start-up in tokamaks

Nuclear fusion is a promising long-term solution to the global energy challenge, offering a safe, abundant and low-carbon energy source[1]. The tokamak is currently the dominant approach toward the development of magnetically confined nuclear fusion[2,3]. Plasma start-up, the foundation of tokamak operation, generally refers to the processes of neutral gas breakdown (plasma generation), formation of toroidal plasma current and closed flux surfaces, further ionization of neutrals (burn-through) and subsequent plasma current ramp-up[4]. Ensuring the robustness and quality of plasma start-up is of great significance, as it exerts a direct and decisive influence on the achievement of target plasmas. It is expected to be challenging in large superconducting, reactor class tokamaks such as ITER, currently under construction in Southern France, due to the low inductive toroidal electric field and the strong plasma-wall interaction (PWI) which occurs as a result of the standard approach used in the majority of current research devices to build-up the plasma in this early phase[5,6].

In the first decades of tokamak research, the majority of devices constructed around the world used largely circular magnetic configurations in which material surfaces defined the last closed flux surface (LCFS). These are referred to as "limiter" configurations since the plasma is bounded or "limited" by a solid component. Although simpler to create, such limiter plasma geometries have the major drawback that impurities released (notably by physical sputtering) at the plasma-wall interaction areas have a high probability of crossing the LCFS into the core where they may seriously impact plasma confinement due to induced radiative power losses and fuel dilution[7]. The situation was radically improved by the development of so-called X-point configurations in which a null is created in the poloidal magnetic field such that open field lines outside the LCFS (referred to as the separatrix in these geometries) can be "diverted" away to material surfaces (divertor targets) more remote from the main plasma chamber. In this way, plasma recycling and PWI are localized away from the core region, leading to much higher confinement performance[8]. Such divertor geometries have now been adopted in all modern research devices and are standard in next step reactor class tokamaks such as ITER[9].

Even in divertor tokamaks, limiter plasmas have not been completely eliminated. In fact, the majority of current devices use a short limiter phase at beginning of each plasma pulse to establish closed flux surfaces, raise the plasma current and prepare the transition to X-point magnetic geometry. In medium to large sized research devices such as EAST, WEST, ASDEX Upgrade or JT-60SA, where plasma currents typically reach several hundred kA or up to ~1 MA at the X-point transition, this early phase is generally of quite short duration (typically $\leq$ 1 s, at most a few seconds)[10,11]. At the ITER scale, where multi-megampere plasmas will be reached at X-point formation, these timescales can be ~10 s[12], due in part to the use of superconducting coils (on which applied voltages must be limited so that current ramp-up rates are reduced) and also as a consequence of settling times for the plasma control system to obtain stable feedback control.

Until the recent 2024 re-baseline[6], ITER's actively cooled first wall (FW) armour was to be manufactured in beryllium (Be). In this case, the impurity influx to the core plasma during the start-up limiter phase was expected to be much less problematic as a consequence of the low Be atomic number. This leads to low radiative losses from the confined plasma and hence has no real impact on the discharge power balance during the low power limiter phases. However, the low radiation loss means high power deposition on FW surfaces at plasma contact, possibly leading to power handling issues in the case of longwave misalignment of the FW components with respect to the tokamak toroidal magnetic field[13,14]. With the switch from Be to the much more reactor relevant tungsten (W) FW armour in the re-baseline, the situation is very different. Sophisticated plasma boundary simulations for ITER[15] and supporting limiter start-up experiments on the all-W EAST, WEST and ASDEX-Upgrade tokamaks[6,16,17] have now demonstrated the clear existence of a self-regulated W sputtering feedback loop which leads to relatively high, but stationary core radiated power losses.

Unfortunately, even if the self-regulated sputtering results in tolerable core W concentrations during limiter start-up, there is a major remaining issue with W main chamber PFCs: the natural oxygen (O) gettering action offered by Be in the previous ITER baseline is now absent and, in the case of elevated core plasma O concentrations during the early plasma current ramp-up phase, "burn-through" to higher currents can be easily compromised by strong radiative losses. This is indeed routinely observed in smaller, all-W research tokamaks[16,17] and is usually mitigated by a wall conditioning technique known as boronization, in which plasma-assisted chemical vapour deposition is used to put down a thin (tens/hundreds of nm) coating of boron (B) on FW surfaces[18,19]. This provides a degree of O gettering and facilitates plasma start-up especially following maintenance periods in which the tokamak vacuum vessel is vented to atmosphere[20–22]. Building on this well-established experience, ITER has decided to install a boronization system[6], but it is very questionable if such an active conditioning process, particularly if it were required on a routine basis to assist start-up, could be considered fusion reactor relevant. The B layer is eroded rapidly in long pulse tokamaks, so that the beneficial effects of the conditioning do not persist for long[19]. For this reason in particular, it would be extremely advantageous if a reliable and robust method could be found to eliminate completely the limiter start-up phase in future reactors since the initial core plasma would be well shielded by the divertor action from even the increased W produced by the presence of O.

It is with these issues in mind that a novel plasma start-up scheme with divertor magnetic geometry only ("DIVO start-up") has recently been developed for the first time on the EAST superconducting tokamak. First results are extremely promising, demonstrating very substantial reductions in core plasma impurity concentrations compared with the standard EAST inboard limiter start-up. The key breakthrough in this work is the formation of a "double-null" divertor geometry with two X-points on the separatrix right after the closure of flux surfaces using an electron cyclotron wave (ECW) generated, non-inductive discharge in a specifically designed poloidal magnetic field configuration (described in detail below). In addition to radically improving the impurity contamination issue, this early phase non-inductive plasma approach would avoid the drawback of the low inductive electric fields in reactor-scale superconducting tokamaks. Once the early plasma is formed, the poloidal field (PF) coil currents are evolved according to designed waveforms, the plasma current increases due to the electron cyclotron current drive (ECCD) and Ohmic current drive (inductive electric field induced by the change of PF coil currents) and the flux surfaces evolve gradually to the final target shape.

## 2. Achievement of divertor geometry in the ECW non-inductive discharge

The key ingredient in the DIVO recipe is the configuration of the poloidal field associated with the initial ECW non-inductive discharge. The characteristics of different field configurations and their experimental realization are discussed below.

The DIVO start-up requires robust formation of closed flux surfaces with X-points and plasma instability control as soon as possible after plasma breakdown. The eddy current in conductive components (principally the vacuum vessel) can deform the desired magnetic field configuration during a conventional inductive plasma start-up, increasing the uncertainty in breakdown time and position and posing challenges for plasma shape control. These issues are considerably alleviated by opting for a non-inductive discharge with ECW pre-ionization and heating. The use of ECW non-inductive discharges is not new and has been systematically investigated in spherical (tight aspect ratio) tokamaks such as CDX-U[23,24], LATE[25–27], TST-2[28,29], SUNIST[30], QUEST[31] and EXL-50[32,33], and also in conventional tokamaks (e.g. DIII-D[24,34], KSTAR[35] and JT60-U[27]). A common finding is the need for an appropriate poloidal magnetic field to ensure the formation of non-inductive toroidal current, an explanation for which is the poloidal field induced asymmetric particle confinement[36].

Compared with conventional quasi-uniform vertical field configuration (VFC) shown in Fig. 1(a), the non-uniform vertical field (stronger at the top and bottom, weaker in the middle) known as a trapped particle configuration (TPC) shown in Fig. 1(b) has been widely used due to the advantage in electron confinement related to the mirror effect. Closed flux surfaces have been obtained with plasma currents at the level of several kA in small spherical tokamaks[23,26]. However, it is difficult to form closed surfaces in large devices even with much larger plasma current[27].

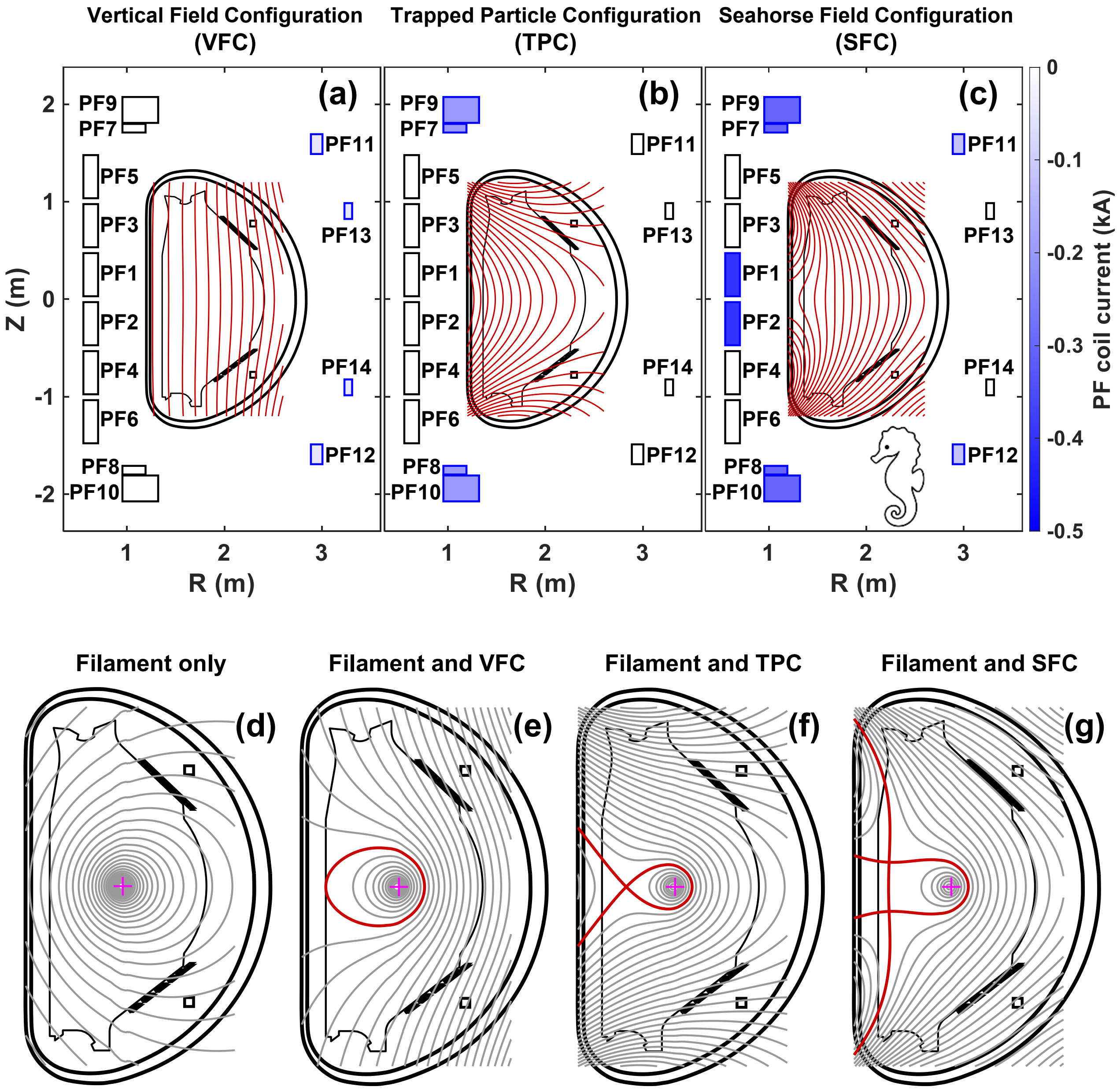

Fig. 1 Upper row shows the vacuum poloidal field configurations with coil currents only: (a) vertical field configuration (VFC), (b) trapped particle configuration (TPC) and (c) seahorse field configuration (SFC). The red contours denote the poloidal magnetic flux, with the corresponding coil currents given by the colour scale. Lower row shows the flux geometries (gray contours) with plasma current simplified by a filament current of 3 kA (marked by plus signs): (d) filament current only, (e) filament current and VFC, (f) filament current and TPC and (g) filament current and SFC. The red lines in (e) - (g) depict the LCFS/separatrix.

To investigate the impact of poloidal field configuration on the flux geometry in the presence of plasma current, a simplified model has been developed in which the plasma current is represented as a filament carrying of 3 kA in the centre of the EAST vacuum vessel (major radius, $R = 1.85$ m and $Z = 0$ m). The corresponding flux geometry is shown in Fig. 1(d). The three different poloidal field configurations in Fig. 1(a-c) including the 3 kA filament are compared in Fig. 1(e)-(g). As shown in Fig. 1(e), the filament current in the VFC usually produces a limiter geometry, even if an X-point can form in a very narrow parametric domain. As a consequence, the VFC variant was excluded as a candidate for DIVO start-up.

In the case of the TPC with the filament current, a divertor geometry is obtained (Fig. 1(f)), with an X-point on the high field side (HFS) midplane. Such a configuration has in fact been achieved on the DIII-D tokamak and so this variant can be considered an option for DIVO start-up. Its feasibility has therefore been tested experimentally on EAST with the result that no closed flux surfaces were observed in the majority of ECW non-inductive discharges. This is because the coil-induced vertical field on the HFS is too large to be cancelled by the plasma current induced vertical field. In the few discharges with an X-point that were obtained, plasma radial position control was challenging due to the horizontal displacement instability[37]. If the latter is not stabilized, a limiter geometry is formed again when the plasma moves outwards and contacts PFCs on the low field side (LFS). These difficulties rule out the TPC as a robust candidate for DIVO start-up.

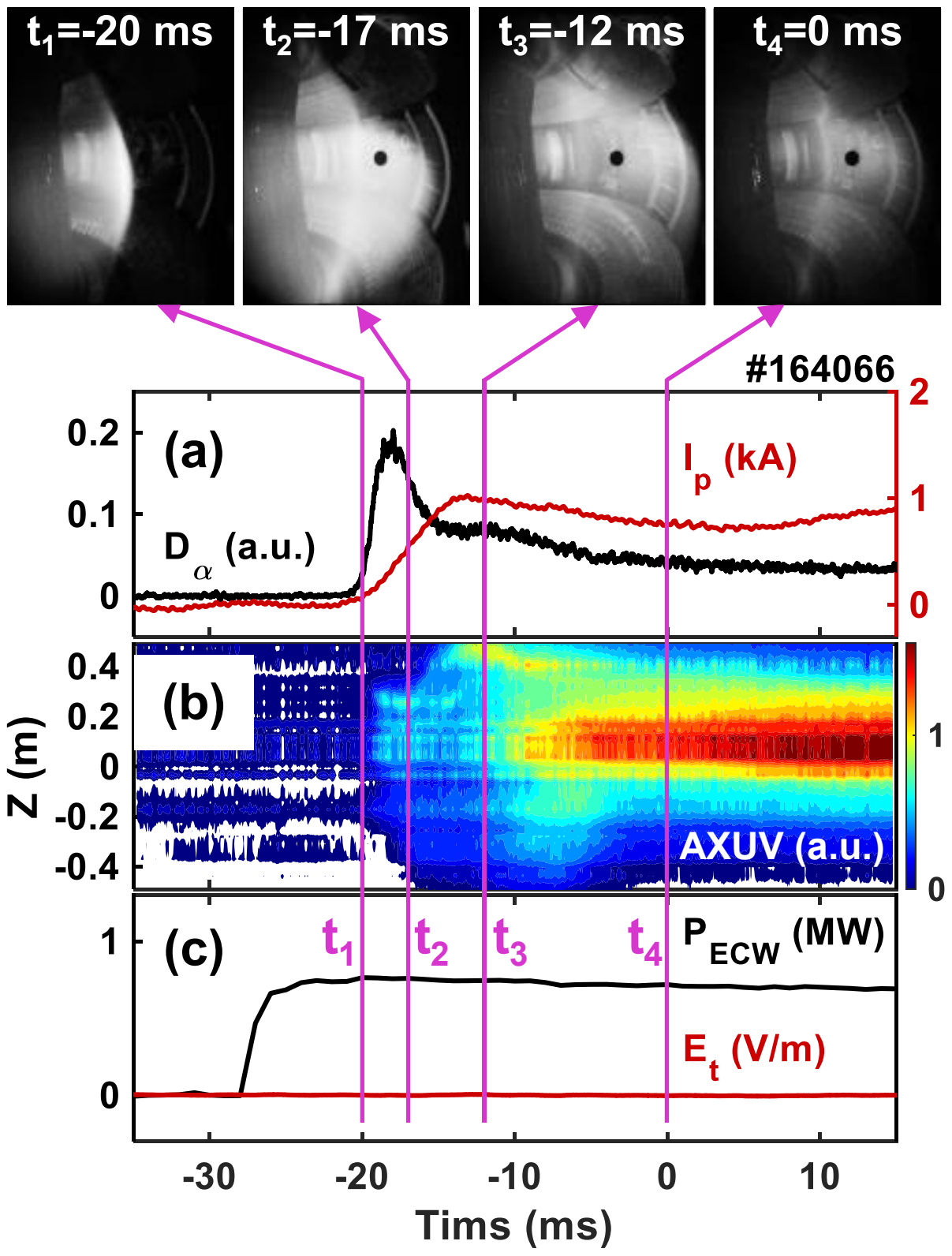

Fig. 2 Example of an ECW non-inductive discharge with the new seahorse-like initial magnetic configuration (SFC) in which the formation process of double-null geometry can be observed. The upper images are visible CCD camera frames at different instants with respect to the time ($t = 0$) at which closed flux surfaces are formed. Lower plots are the temporal evolution of (a) $D_\alpha$ emission (black) and plasma current, $I_p$ (red); (b) spatial-temporal evolution of absolute extreme ultraviolet (AXUV) emission where the Z-coordinate represents the vertical positions of the intersection between each line-of-sight (LoS) of the AXUV channels and $R = 1.85$ m (major radius of the vacuum vessel centre); (c) ECW power (black) and inductive toroidal electric field (red).

The poloidal magnetic field structure shown in Fig. 1(c) and with the plasma current filament added in Fig. 1(g), is a double null divertor geometry with two X-points on the separatrix and located inside the vacuum vessel. It is combination of a barrel-like field (see for example the vacuum field in Extended Data Fig. 1(b)) around the centre of the vacuum vessel and a large TPC (mirror-like) field across the whole poloidal cross-section. Inspired by the shape of this combined poloidal magnetic field geometry, this configuration is referred to here as the Seahorse Field Configuration (SFC) – see insert in Fig. 1(c). It has a number of important advantages: 1) high electron confinement due to the mirror effect; 2) considerably eased formation of closed flux surfaces due to the weak HFS vertical field and; 3) weak horizontal and vertical displacement instability due to the low poloidal field gradient in the centre of the vacuum vessel.

The SFC performance has also been assessed in EAST experiments. Fig. 2 provides an example of an ECW non-inductive discharge (inductive toroidal electric field $E_{\mathrm{t}}{\sim}0$) with power $P_{\mathrm{ECW}} = 0.7\ \mathrm{MW}$ in which the pronounced increase in deuterium Balmer-α line emission ($\mathrm{D}_{\alpha}$)[38] and the bright region of visible light in the CCD image[39] at the ECW resonance layer at time $t_1$ indicate the breakdown of neutral gas. A vertical electric field subsequently forms due to charge separation arising from the opposite $B \times \nabla B$ drifts of electrons and ions. The radial expansion of plasma at time $t_2$ is mainly attributed to the resulting $E \times B$ drift. At later times (from $t_3$ to $t_4$), both the bright region in the CCD camera images and the absolute extreme ultraviolet (AXUV) emission[40] change from a more diffuse to a more concentrated state, indicating the formation of closed flux surfaces. The numerically calculated X-points shown in Fig. 1(g) can be observed in the CCD image at time $t_4$ in Fig. 2.

## 3. Full demonstration of the DIVO start-up on EAST

The entire process of DIVO start-up has been demonstrated on EAST as shown in Fig. 3. Initial pre-magnetization of the PF coils begins at $t = -5.5$ s, followed by establishment of the SFC configuration from $t = -0.5\ \mathrm{s}$ to $t = 0\ \mathrm{s}$ during which the prefill gas is injected. The 0.7 MW of ECW power was switched on at $t = -0.028\ \mathrm{s}$ for pre-ionization and subsequent auxiliary heating. The double-null geometry (Stage 1 in Extended Data Fig. 1(a)) was achieved at the end of the ECW non-inductive discharge (from $t = -0.028\ \mathrm{s}$ to $t = 0$ s). After this point, the PF coil currents evolve following the waveforms shown in Extended Data Fig. 2(c), implying the application of inductive loop voltage.

The evolution of both plasma current and plasma shape in the first experimental tests was found to be the same as that predicted before $t = 0.1$ s, after which the growth of a vertical displacement event (VDE) led to discharge termination. To remedy this, pure feedforward programming control was switched at 0.1 s to $RZI_p$ control (a plasma control scheme considering the horizontal position $R$ and vertical position $Z$ of the plasma centre, and plasma current $I_{\mathrm{p}}$ as controlling variables[10]) including VDE feedback control. In the $RZI_p$ control phase, the pre-designed PF coil current waveforms still play a role in plasma shape control as they did previously in pure feedforward control. After $t = 0.4\ \mathrm{s}$ with $I_{\mathrm{p}} > 100$ kA, accurate magnetic equilibrium reconstruction using the EFIT code[41] based on magnetic measurements becomes available. From this point it is then possible to use ISO-flux control (a plasma control scheme which requires real time equilibrium reconstruction and achieves precise plasma shape control by regulating the flux at each control point to match that at a reference point, typically the X-point[42]) and the plasma current can be ramped up in the usual way, successfully demonstrating the full DIVO start-up scheme. The robustness of the DIVO start-up has also been verified on EAST with varying prefill gas pressure, impurity level and inductive electric field.

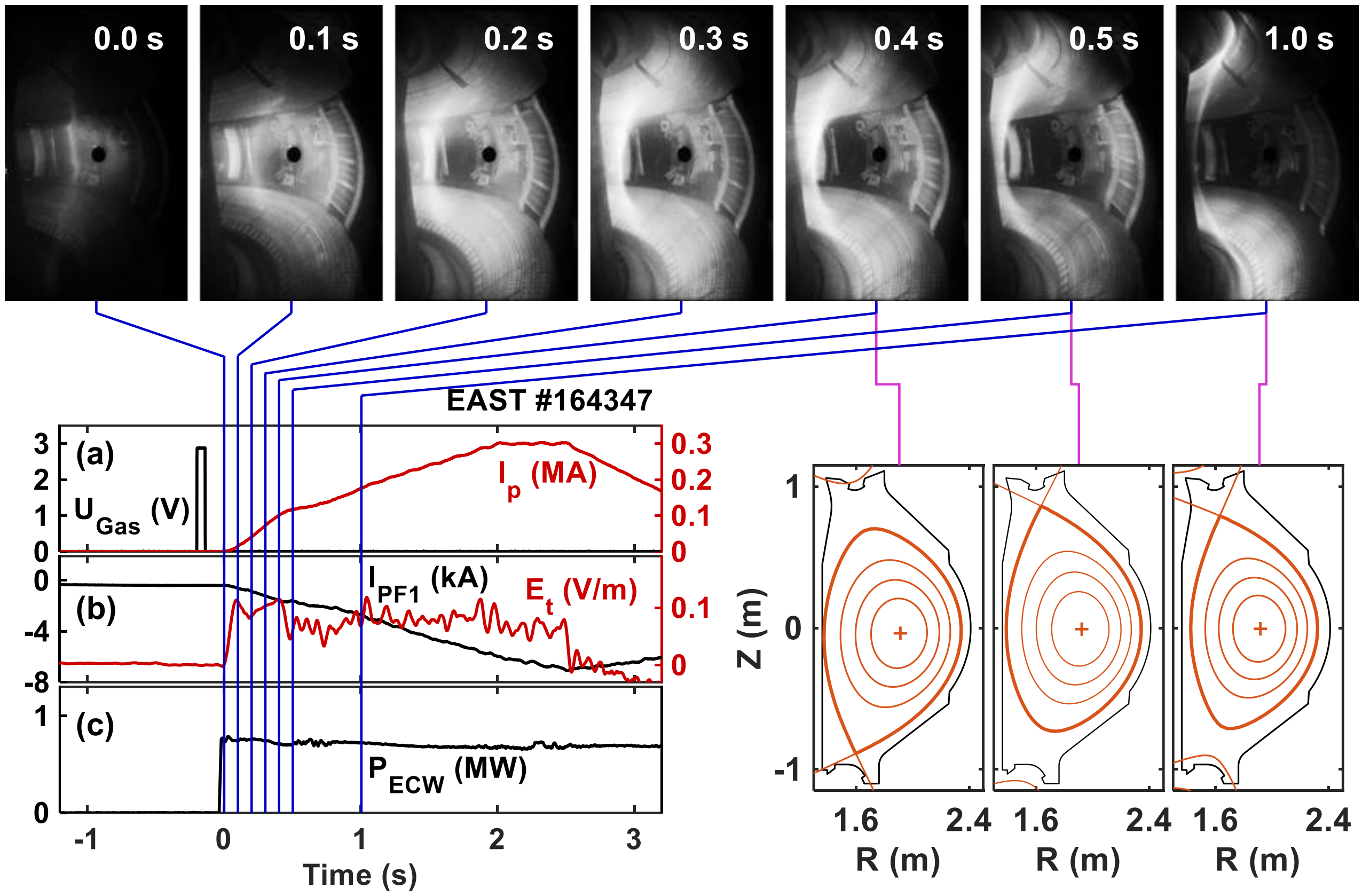

Fig. 3. Demonstration of the DIVO start-up on EAST. (a) control voltage for the prefill gas $U_{\mathrm{Gas}}$ and $I_{\mathrm{p}}$; (b) current waveform of coil PF1 $I_{\mathrm{PF1}}$ (see Fig. 1 for the PF coil numbering) and $E_{\mathrm{t}}$ (PF1 is a part of the central solenoid whose current change contributes most to the induced toroidal electric field); (c) ECW power. Visible light CCD camera images at selected times are shown in the upper row, while the flux contours at lower right are equilibrium reconstruction results based on magnetic measurements at 0.4 s, 0.5 s and 1 s showing the development of first lower single null and then the final upper single null shape which remains unchanged through the rest of the $I_{\mathrm{p}}$ ramp-up and flattop. Note the much larger time extent of the x-axis compared to Fig. 2 which focusses only on the early initiation phase.

## 4. Decrease of impurity emission with DIVO start-up

As discussed in Section 1, a major potential benefit of the DIVO start-up scheme is the avoidance of limiter contact during the early current ramp-up. This is a particular issue in all-W devices, especially if concentrations of low to medium atomic number impurities such as O are present as adsorbed species or bound in oxides on the PFC surfaces. These impurities can then be very efficiently recycled into the confined plasma and lead to intense radiative losses which threatens successful burn-through. Starting the discharge directly in divertor configuration should strongly mitigate this effect, mainly due to the much lower impurity sources compared to those in limiter configuration. In addition, the impurities released in the divertor strike points are much more efficiently screened by the X-point geometry.

Figure 4 illustrates this by comparing a pair of consecutive discharges with first a conventional start-up (HFS limiter configuration) and then the new DIVO scheme. A key point to note concerning this pair of discharges is that they were performed 4 plasma operational days away from the last vessel boronization, which corresponds to >3200 s of total discharge duration, considerably longer than the ~1700 s typical lifetime of B coatings on EAST[19]. Thus, the discharges compared here do not benefit from the very clean conditions with low plasma O content that are obtained immediately after a new B layer is deposited. This provides a better opportunity to assess the true benefits of the new limiter-free start-up scheme.

In both discharges, the prefill gas waveforms and ECW configuration, including timing, injection angle and power were identical. Since the plasma is limited on the HFS FW (the central column) in the conventional start-up and has divertor strike points on the HFS in the SFC configuration, Fig. 4 provides time traces of visible impurity emissions from this region. Moreover, given that the EAST inner wall is armoured with both W and graphite (C) tiles, line radiation from neutral W atoms (WI), $C^{2+}$ ions (CIII) and $O^{1+}$ ions (OII) is included. Evidently, before $t = 1.1$ s all impurity emission intensities during the DIVO start-up (shot #165124) were much lower than those in the limiter phase of the conventional scenario (shot #165124). Beyond 1.1 s, when the plasma has transited to the divertor configuration in the conventional start-up, the line intensities are very low and similar in both schemes. With the exception of the very early phase (for $t < \sim 0.3$ s), the plasma current ramp-up waveforms are practically identical in both discharges. These are powerful indicators that the direct divertor start-up scheme is greatly beneficial in reducing both the low Z impurity content and the W source in the critical early phases of plasma current formation.

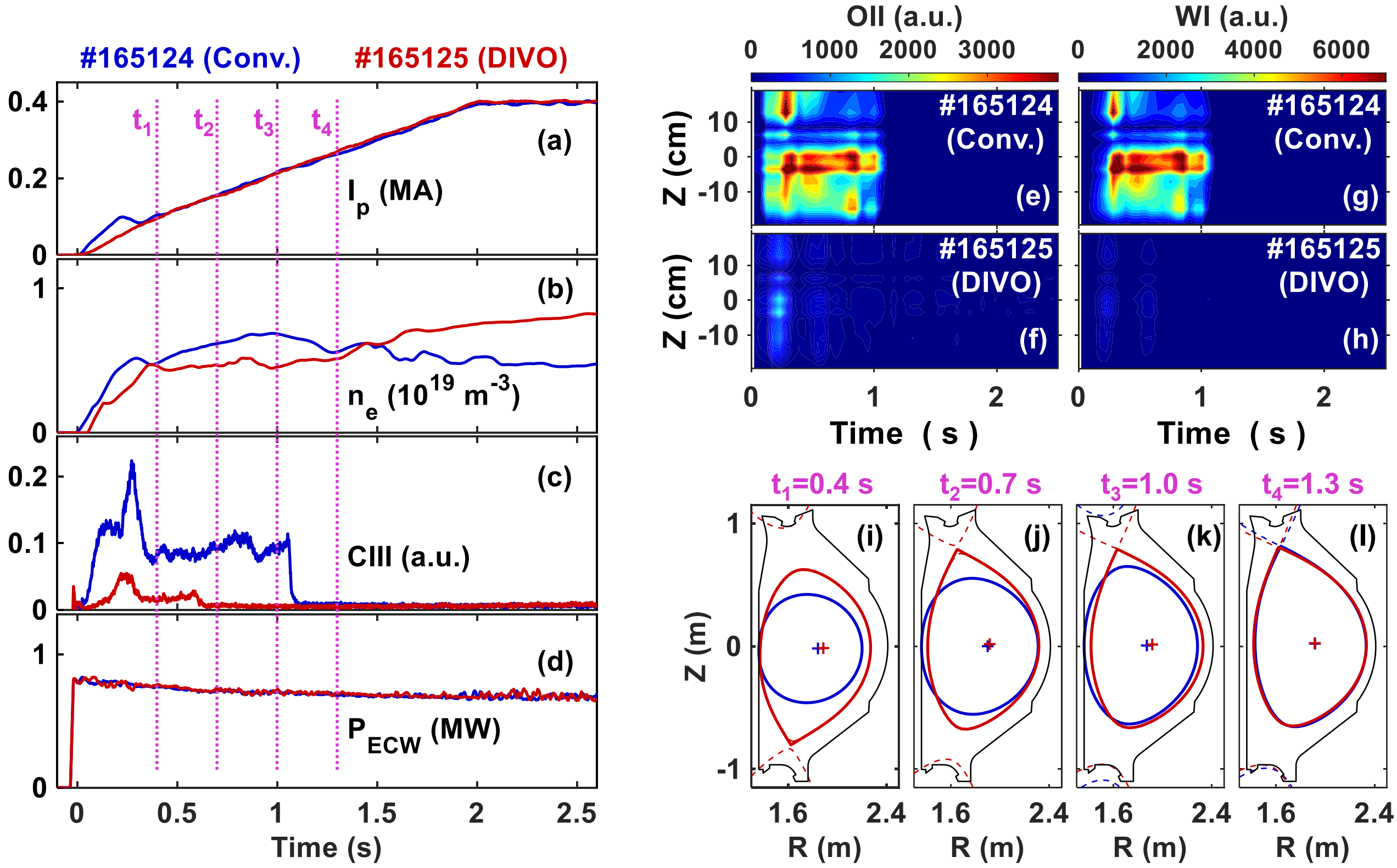

Fig. 4. Comparison of (a) plasma current, (b) line-averaged electron density, (c) CIII line emission from $C^{2+}$ ions, (d) ECW power $P_{ECW}$, (e)-(f) OII line emission from $O^{1+}$ ions, (g)-(h) WI line emission from neutral W atoms, and (i)-(l) plasma shapes calculated by the EFIT equilibrium reconstruction code at the four time points (marked by the dotted lines in the left column) in two consecutive discharges with DIVO start-up (#165125) and conventional start-up (#165124). The plasma shape in shot #165124 transits from limiter to divertor geometry at $t = 1.1$ s. The Z-coordinate in (e)-(h) represents the vertical positions of the intersection between each LoS of the optical diagnostic and the FW on the HFS.

## 5. Conclusion and discussion

DIVO start-up, a novel plasma start-up scheme with divertor geometry only has been proposed and recently demonstrated for the first time on the EAST tokamak. The new scheme offers several advantages over the conventional approach presently used throughout the world tokamak community, and planned in the ITER reactor class device, in which the initial plasma current ramp phase is formed in a limiter configuration. It employs a non-inductive discharge with electron cyclotron wave pre-ionization and heating and, crucially, significantly reduces impurity influxes through the immediate formation of divertor configuration. This strongly beneficial impact on plasma contamination compared to the standard start-up scenario has been clearly observed in the EAST experiments and firmly establishes this new scheme as an alternative to conventional limiter ramp-up. The reduction in the initial impurity level is particularly meaningful when the available heating power is insufficient, as is often the case during the commissioning phase of a new device. Moreover, the DIVO start-up can reduce the heat load on plasma facing components by avoiding the direct contact inherent in the conventional limiter start-up scheme, thereby offering the possibility of reducing the required heat-handling capability of plasma facing components and, consequently, the manufacturing cost of tokamak reactors.

Further efforts are nevertheless required to substantiate the claim of this approach for application in next-step superconducting devices with all-tungsten plasma-facing components. In particular, the dependence of the required ECW power on machine size and the plasma control capability in reactor class devices, in which the plasma is much further from the poloidal field coils because of the tritium breeding blanket, must be established. Studies across different devices are planned to consolidate the DIVO scheme as a viable start-up alternative for next-step tokamaks.

# Methods

## Experimental methodology

The Experimental Advanced Superconducting Tokamak (EAST) is a fully superconducting, full-tungsten tokamak, located at the Institute of Plasma Physics, Chinese Academy of Sciences (ASIPP), in Hefei, China[43] . Its main design parameters include a major radius $R = 1.85$ m, minor radius $a = 0.45$ m, and plasma current in the range $0.3 - 1$ MA. The ECW system uses 140 GHz gyrotrons[44,45] and employs a second harmonic X-mode (X2) scheme. The toroidal ECW launch angle in this work was set to $200°$ ($20°$ oblique in the co-current direction). The toroidal magnetic field at $R = 1.85$ m was 2.48 T, with the X2 resonance layer located at $R{\sim}1.83$ m. The diagnostics used include the visible spectroscopy systems[38,46] monitoring the line emissions from C, O and W impurities, the absolute extreme ultraviolet (AXUV) emission diagnostic[40] , the interferometer[47] measuring the line-integrated electron density, a fast visible CCD camera[39] providing plasma images, and the magnetic measurement system[48] for equilibrium reconstruction.

## Design of the PF coil current waveforms for DIVO start-up

The formation of double-null geometry in the ECW non-inductive discharge with SFC is the first stage of DIVO start-up. Considering that the plasma current at this stage (<2 kA as shown in Fig. 2) is too low to reconstruct an equilibrium based on magnetic measurements, it is necessary to adopt feedforward programming control until feedback control becomes possible. The numerical calculation process of PF coil current waveforms in the feedforward control phase of DIVO start-up is introduced below.

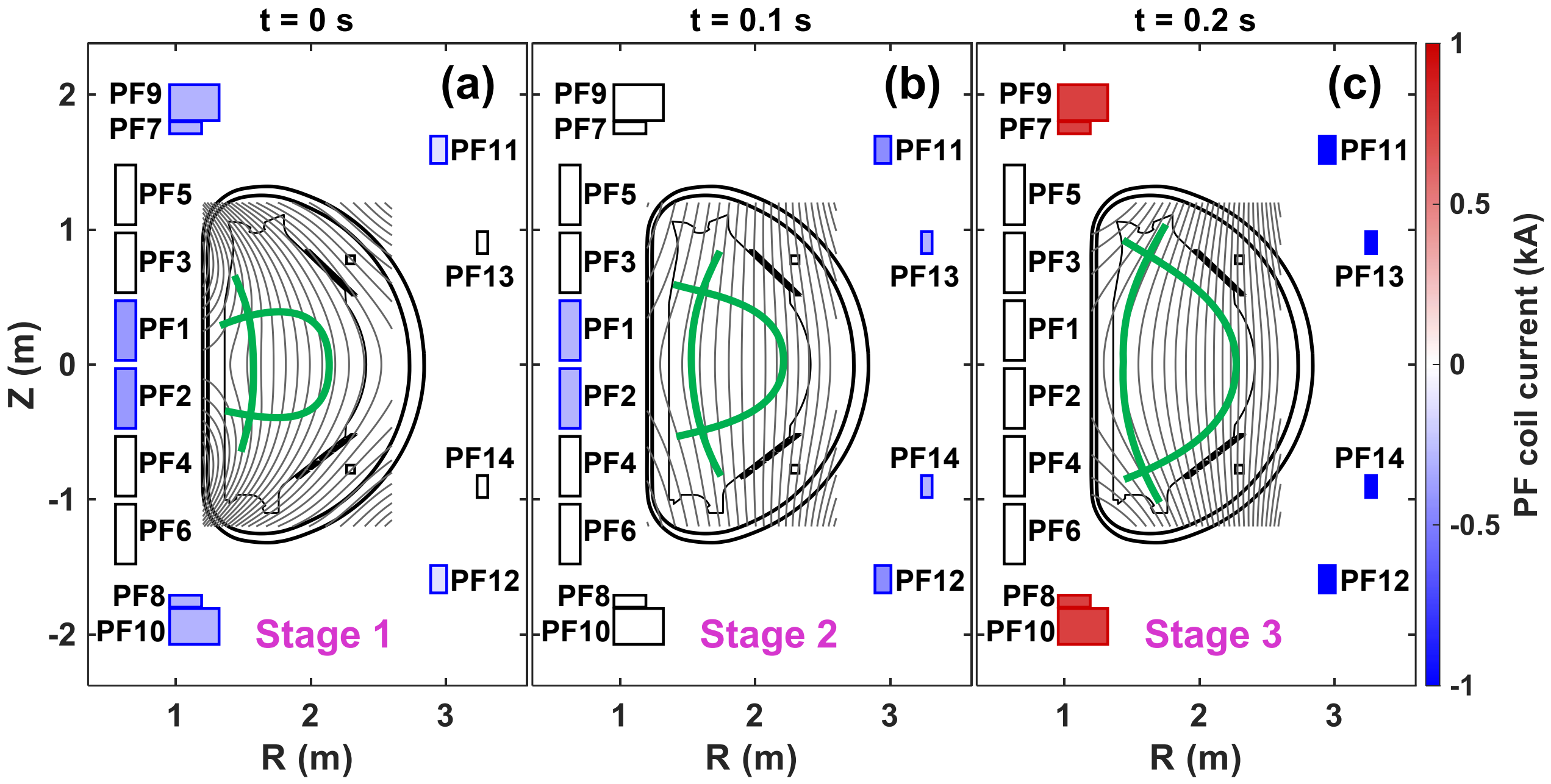

Extended Data Fig. 1| Vacuum field configurations (grey contours) and corresponding plasma shapes (green lines) for DIVO start-up at (a) stage 1, (b) stage 2 and (c) stage 3.

The first step is to consider plasma shape control. Extended Data Fig. 1 shows three typical stages of the DIVO start-up, illustrating the evolution of the vacuum field configuration (grey contours) and the corresponding plasma shape (green lines). The vacuum field at Stage 1 is the SFC in Fig. 1(b) of the main text. The plasma shape is the double-null geometry shown in Fig. 1(g) and in the CCD image at time $t_4$ in Fig. 2. Stage 2 is a transition state. Stage 3 corresponds to a double-null geometry with X-point positions close to those of the final target shape. The transition times between adjacent stages are designed to be 0.1 s according to the plasma current ramp rate and the rate of change of PF coil currents in routine EAST operation. The corresponding PF current waveforms for plasma shape control $I_{\mathrm{PF,S}}$ shown in Extended Data Fig. 2(a) are determined by spline interpolation based on the three stages.

Once the shape control has been defined, the next step is to design the PF coil currents for plasma current control. The inductive toroidal electric field $E_{\mathrm{t}}$ induced by the evolution of PF coil currents was calculated by the Green's function method. Based on experimental data obtained during routine EAST operation, $E_{\mathrm{t,T}}$, the target toroidal electric field at the centre of vacuum vessel for DIVO start-up was set to 0.22 V/m corresponding to a loop voltage of ~2.5 V with a rise time of ~0.04 s (black line in Extended Data Fig. 2(b)). In addition, $E_{\mathrm{t,S}}$, the toroidal electric field contributed by the evolution of PF coil currents for shape control (red line in Extended Data Fig. 2(b)) was calculated according to the waveforms shown in Extended Data Fig. 2(a). An extra toroidal electric field $E_{\mathrm{t,C}} = E_{\mathrm{t,T}} - E_{\mathrm{t,S}}$ (blue line in Extended Data Fig. 2(b)) was required for plasma current control only with as little influence on plasma shape as possible. To meet this requirement, PF coil currents should change according to the ratios of the null field configuration. The corresponding waveforms for plasma current control $I_{\mathrm{PF,S}}$ are shown in Extended Data Fig. 2(c).

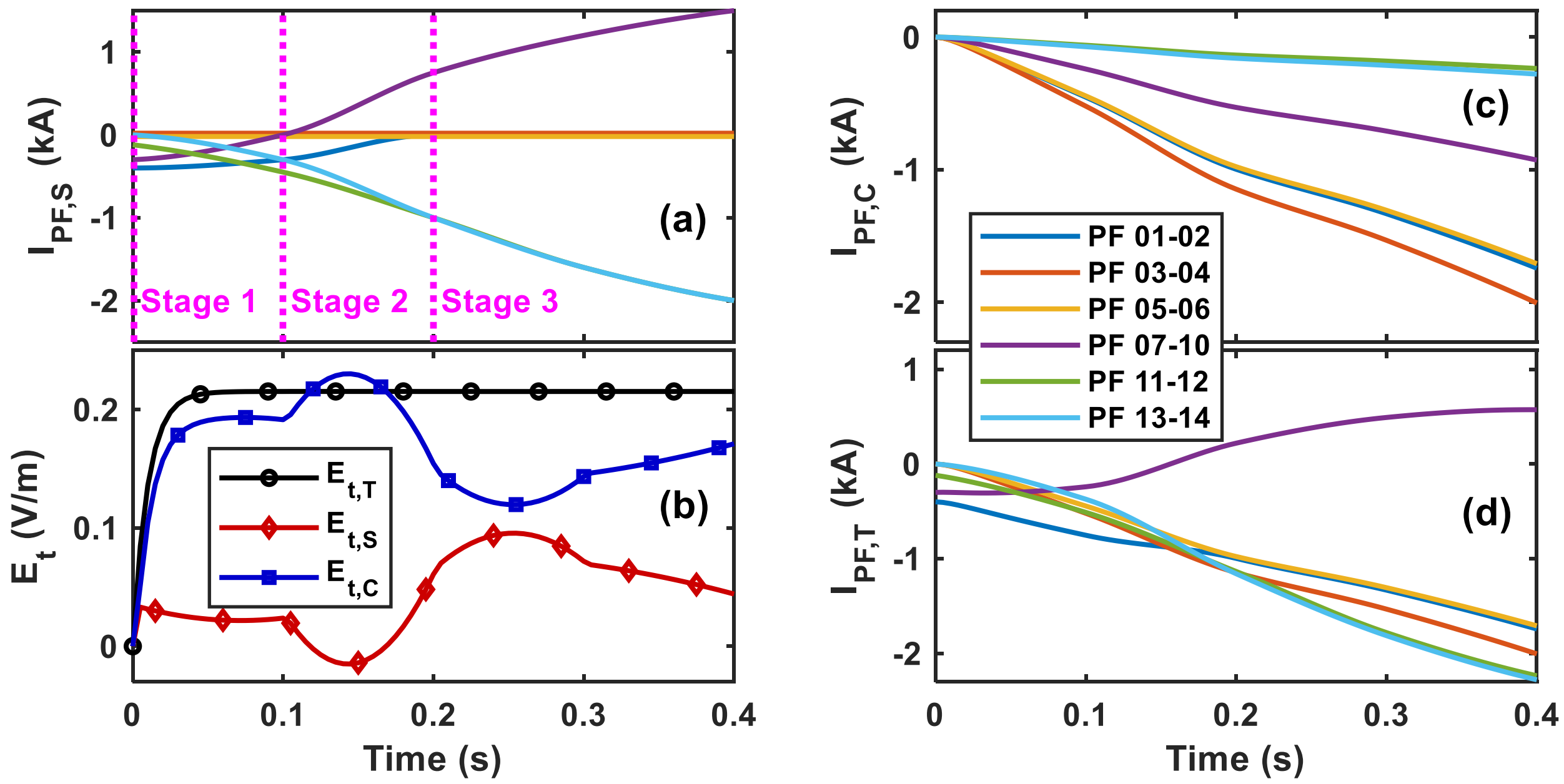

Extended Data Fig. 2| (a) PF current waveforms for plasma shape control $I_{\mathrm{PF,S}}$ where the dashed lines indicate the three stages in Extended Data Fig. 1(a)-(c); (b) target toroidal electric field at the centre of vacuum vessel for DIVO start-up ($E_{\mathrm{t,T}}$), toroidal electric field contributed by the evolution of PF coil currents for shape control ($E_{\mathrm{t,S}}$) and extra toroidal electric field ($E_{\mathrm{t,C}} = E_{\mathrm{t,T}} - E_{\mathrm{t,S}}$) required for plasma current control only; (c) PF coil current waveforms for plasma current control $I_{\mathrm{PF,C}}$ corresponding to the extra loop voltage $E_{\mathrm{t,C}}$; (c) PF coil current waveforms in total $I_{\mathrm{PF,T}}$ finally used in experiments meeting the requirements of both plasma shape control and current control.

Finaly, a summation of the waveforms in Extended Data Fig. 2(a) and in Extended Data Fig. 2(c) yields the waveforms of PF coil current in total $I_{\mathrm{PF,T}}$ actually used in experiments (Extended Data Fig. 2(d)), which meet the requirements of both plasma shape and current control. It should be noted that, for the purposes of simplification, the impact of eddy currents in conducting components such as the vacuum vessel on plasma shape and current control was not considered in the calculation. The experimental results shown in the main text demonstrate that the neglect of eddy currents is justified.

# Acknowledgements

We thank the staff members at the EAST (Experimental Advanced Superconducting Tokamak) facility in Hefei (https://cstr.cn/31130.02.EAST) for their technical support and assistance in data collection and analysis. This work was supported by the National MCF Energy R&D Program (2025YFE03000100 and 2024YFE03010000), the National Natural Science Foundation of China under Contract Nos. 12475230 and 12425509, and the Anhui Provincial Major Science and Technology Project 2023z020004. The views and opinions expressed herein do not necessarily reflect those of the ITER Organization.